\documentclass[9pt,conference]{IEEEtran}
\IEEEoverridecommandlockouts

\usepackage{amsmath,amsfonts,amssymb,amsthm,enumitem,graphicx}
\usepackage{xcolor}        
\usepackage{color}
\usepackage{float}
\usepackage{caption}
\usepackage{booktabs}
\usepackage{multirow}
\usepackage{mathtools}
\usepackage{epstopdf}
\usepackage{subfigure}
\usepackage{hyperref}
\usepackage{setspace}
\usepackage{cite}
\usepackage{rotating}

\usepackage[mathcal]{eucal}
\usepackage[bb=pazo]{mathalpha}

\usepackage{mathtools}

\usepackage{tikz}
\usetikzlibrary{positioning}

\makeatletter
\g@addto@macro\normalsize{%
  \setlength\intextsep{1ex}
  \setlength\textfloatsep{1ex}
  \setlength\abovecaptionskip{1ex}
  \setlength\belowcaptionskip{1ex}
}\makeatother

\DeclareMathOperator*{\argminopt}{argmin}
\DeclareMathOperator*{\minopt}{min}

\DeclareMathOperator*{\proxopt}{prox}

\usepackage{gensymb}
\usepackage{amsmath}
\usepackage{graphicx}
\usepackage{algorithm}
\usepackage{algorithmic}
\usepackage[ruled,vlined,algo2e]{algorithm2e}

\makeatletter
\g@addto@macro\normalsize{%
  \setlength\abovedisplayskip{1ex}
  \setlength\belowdisplayskip{1ex}
  \setlength\abovedisplayshortskip{1ex}
  \setlength\belowdisplayshortskip{1ex}
}\makeatother

\newcommand{\subf}[2]{%
  {\small\begin{tabular}[t]{@{}c@{}}
  #1\\#2
  \end{tabular}}%
}

\makeatletter
\def\old@comma{,}
\catcode`\,=13
\def,{%
  \ifmmode%
    \old@comma\discretionary{}{}{}%
  \else%
    \old@comma%
  \fi%
}
\makeatother

\def\BibTeX{{\rm B\kern-.05em{\sc i\kern-.025em b}\kern-.08em
    T\kern-.1667em\lower.7ex\hbox{E}\kern-.125emX}}

\begin{document}

\captionsetup[algorithm]{labelformat=simple, labelsep=colon}

\title{Online Proximal ADMM for Graph Learning\\ from Streaming Smooth Signals
\thanks{This research was supported by the IEEE SPS ME-UYR program.}
}

\author{\IEEEauthorblockN{Hector Chahuara${}^{\dagger}$ and Gonzalo Mateos${}^{\ddagger}$}
\IEEEauthorblockA{${}^{\dagger}$Department of Electrical Engineering, Pontificia Universidad Cat\'olica del Per\'u, Lima, Per\'u\\$^{\ddagger}$Department of Electrical and Computer Engineering, University of Rochester, Rochester, NY, USA}}

\maketitle

\begin{abstract}
Graph signal processing deals with algorithms and signal representations that leverage graph structures for multivariate data analysis. Often said graph topology is not readily available and may be time-varying, hence (dynamic) graph structure learning from nodal (e.g., sensor) observations becomes a critical first step. In this paper, we develop a novel algorithm for \emph{online} graph learning using observation streams, assumed to be smooth on the latent graph. Unlike batch algorithms for topology identification from smooth signals, our modus operandi is to process graph signals sequentially and thus keep memory and computational costs in check. To solve the resulting smoothness-regularized, time-varying inverse problem, we develop online and lightweight iterations built upon the proximal variant of the alternating direction method of multipliers (ADMM), well known for its fast convergence in batch settings. The proximal term in the topology updates seamlessly implements a temporal-variation regularization, and we argue the online procedure exhibits sublinear static regret under some simplifying assumptions. {Reproducible} experiments with synthetic and real graphs demonstrate the effectiveness of our method in adapting to streaming signals and tracking slowly-varying network connectivity. The proposed approach also exhibits better tracking performance (in terms of suboptimality), when compared to state-of-the-art online graph learning baselines.
\end{abstract}

\begin{IEEEkeywords}
Topology identification, dynamic network tracking, proximal ADMM, smooth signal, regret analysis
\end{IEEEkeywords}

\section{Introduction}
\label{sec:introduction}

Graphs serve as powerful abstractions of the underlying relational structure among data entities (e.g., sensors, brain regions), thus becoming a viable modeling tool for a host of signal processing and machine learning applications. Accordingly, there has been growing interest in describing multivariate complex data, such as social networks, neural activity time series, and urban traffic flows, as graph signals; see e.g.~\cite{gsp2018tutorial}. However, this valuable (latent) graph structure may not be explicitly available, and so it needs to be estimated from nodal data before any meaningful downstream analysis can be conducted~\cite{Mateos2019,Dong2019}. Networked systems (as the examples above) are in constant flux, with increasing size and complexity, so there is a need for efficient topology inference algorithms that can process data streams generated by \emph{dynamic} networks~\cite{Giannakis2018,emiliano2020}.\vspace{3pt}

\noindent \textbf{Related work.} The inverse problem of searching for a graph (encoded by e.g., an adjacency or Laplacia matrix) that in some sense best describes a given multivariate dataset is known as network topology inference or graph structure learning. Optimality criteria and constraints often arise from statistical priors, physical principles, interpretability, or downstream task goals, which result in models that tie the observations to the latent graph. 
These models can be probabilistic~\cite{Friedman2008Glasso,Pavez2018,vinicius2021graphical}, assume graph signals are stationary or generated by network diffusion~\cite{segarra2017spectralTemplates,pasdeloup2018tsipn,dorina2017diffusion,wasserman2023GDNtmlr,madeline2024joint}, or else they are smooth w.r.t. to the graph \cite{Dong2016,Kalofolias2016,Saboksayr2021-accelerated,Wang2023,wasserman2024DPGtmlr}. In this work, we rely on the latter class.



Specifically, graph learning from smooth signals is formulated as a non-smooth strongly convex optimization problem, and several \emph{batch} solvers have been developed to date. For instance, a scalable primal-dual algorithm was proposed in \cite{Kalofolias2016}, while alternating direction method of multipliers (ADMM) iterations were put forth in~\cite{Wang2021}. An accelerated dual-domain proximal gradient algorithm (DPG) was studied in~\cite{Saboksayr2021-accelerated}, with the first convergence rate results for this problem. The best rates to date are reported in~\cite{Wang2023}, for a proximal ADMM (PADMM) algorithm developed for time-varying graph learning. All these approaches operate in batch mode, they are non-recursive, and hence their computational complexity and memory storage grow linearly with time when used to estimate dynamic network topologies. 

Recognizing these challenges, \emph{online} graph learning methods have been recently developed to process data sequentially-in-time. The online proximal gradient (PG)~\cite{Saboksayr2021} and DPG \cite{Saboksayr2023} approaches can track slowly-varying dynamic graph topologies from streams of smooth signal observations. 
We would be remiss not to mention other online graph learning methods that rely on assumptions other than signal smoothness. For instance, \cite{Vlaski2018} introduces an online algorithm that uses observations from a Laplacian-based continuous-time graph process, while \cite{Shafipour2020} leverages graph signal stationarity. An encompassing (model-independent) framework that exploits a prediction-correction strategy was proposed in~\cite{natali2022online}.\vspace{3pt}

\noindent \textbf{Proposed approach and contributions.} In this paper, our main contribution is to develop a novel online algorithm to track the topology of (potentially dynamic) undirected graphs using streaming smooth signals. Our method builds on PADMM~\cite{Beck2017}, whose provably fast local convergence properties have been well documented in a batch graph learning setting~\cite{Wang2023}. Our technical contributions are significant for several reasons. The proposed online version of PADMM (henceforth dubbed OPADMM): (i) incorporates proximal terms in the topology updates which effectively serve as temporal-variation regularizers; (ii) yields entrywise separable and lightweight updates, with time-independent computational cost and memory footprint; and (iii) exhibits sublinear static regret under simplifying assumptions. Numerical tests with synthetic and real-world graphs demonstrate that PADMM outperforms state-of-the-art online graph learning baselines for this problem, and can seamlessly track slowly-varying dynamic network topologies. {In support of reproducible research, we share the code used to generate all figures in this paper.}

\section{Preliminaries and Problem Formulation}
\label{sec:theoretical-background}

\subsection{Graph-theoretic notions and problem statement}

Consider an unknown undirected graph denoted as $\mathcal{G}\left(\mathcal{V},\mathcal{E},\mathbf{W}\right)$, with vertices $\mathcal{V} = \left\{1,\dots,n\right\}$, edges $\mathcal{E} \subseteq \mathcal{V} \times \mathcal{V}$, and the symmetric adjacency matrix $\mathbf{W} \in \mathbb{R}_{+}^{n \times n}$ that collects the edge weights with $\mathbf{W}_{i,j}=0$ for $(i,j) \notin \mathcal{E}$. Also, $\mathbf{W}_{i,i}=0$, $\forall i \in \mathcal{V}$, since we exclude self-loops. The graph Laplacian is $\mathbf{L} = \textrm{diag}\left(\mathbf{d}\right)-\mathbf{W}$, where $\mathbf{d} = \mathbf{W}\mathbf{1}_n$ represents the vector of nodal degrees

Along with the graph $\mathcal{G}\left(\mathcal{V},\mathcal{E},\mathbf{W}\right)$, the graph signal observations are denoted as $\mathbf{x} \in \mathbb{R}^{n}$, where $x_{i}$ is the value measured at node $i \in \mathcal{V}$. The Dirichlet energy or total variation (TV) quantifies the smoothness of graph signals supported on $\mathcal{G}$ \cite{Zhou2004}, and is given by

\begin{equation}
\label{eq:tv-definition}
\text{TV}\left(\mathbf{x}\right) \coloneqq \mathbf{x}^\top\mathbf{L}\mathbf{x} = \frac{1}{2}\sum_{i \neq j}W_{ij}\left(\mathbf{x}_{i} - \mathbf{x}_{j}\right)^{2},
\end{equation}

\noindent where $\text{TV}\left(\mathbf{x}\right) \in \left[0,\lambda_{\text{max}}\right]$, where $\lambda_{\text{max}}$ is the largest eigenvalue of the positive semidefinite (PSD) matrix $\mathbf{L}$. We say a graph signal is smooth (or low-pass bandlimited) if it has a small total variation. 

We have all the ingredients to state the topology inference problem. Our goal is to learn an undirected graph $\mathcal{G}\left(\mathcal{V},\mathcal{E},\mathbf{W}\right)$ from a stream of graph signal measurements $\mathbf{X} \coloneqq \{\mathbf{x}^{(k)}\}$ that are smooth on $\mathcal{G}$. We also consider tracking dynamic networks with slowly time-varying weight matrix $\mathbf{W}^{(k)}$, $k = 1,2,\dots$ ($\mathcal{V}$ remains fixed throughout).

\subsection{Batch topology identification from smooth signals}

Consider a batch of $p$ graph signals $\mathbf{x}^{(k)}$, $k = 1,\dots,p$  collected as columns of the data matrix $\mathbf{X} = [\mathbf{x}^{(1)},\ldots,\mathbf{x}^{(p)}] \in \mathbb{R}^{n \times p}$. Then form the vertex pairwise dissimilarity matrix $\mathbf{Z} \in \mathbb{R}_{+}^{n \times n}$, with entries $Z_{ij} = \left\|\mathbf{X}_{i,:}-\mathbf{X}_{j,:}\right\|_{2}^{2}$, $(i,j) \in \mathcal{V}$. With these definitions, the aggregate TV measure over $\mathbf{X}$ can be expressed as
\begin{equation}
\label{eq:smoothness-measure}
\sum_{k=1}^{p}\text{TV}\left(\mathbf{x}^{(k)}\right) = \text{tr}\left(\mathbf{X}^\top\mathbf{L}\mathbf{X}\right) = \frac{1}{2}\left\|\mathbf{Z}\odot\mathbf{W}\right\|_{1,1},
\end{equation}
where $\odot$ stands for the Hadamard product and $\|\cdot\|_{1,1}$ denotes the entrywise matrix $\ell_1$ norm. Identity \eqref{eq:smoothness-measure} shows that minimizing the aggregate TV measure w.r.t. $\mathbf{W}$ boils down to preferentially dropping edges $(i,j)$ with higher pairwise nodal dissimilarities $Z_{i,j}$, thus sparsifying $\mathcal{E}$. This link between signal smoothness and edge sparsity was first recognized in~\cite{Kalofolias2016}, which advocates graph structure learning by solving the following convex inverse problem 
\begin{equation}
\label{eq:graph-learning-problem}
    \begin{split}
    &\minopt_{\mathbf{W} \in \mathbb{R}^{n \times n}}\enskip \left\|\mathbf{Z}\odot\mathbf{W}\right\|_{1,1} -\alpha\mathbf{1}_n^\top\log\left(\mathbf{W}\mathbf{1}_n\right) + \frac{\beta}{2}\left\|\mathbf{W}\right\|_{F}^{2}\\ & \text{subject to} \enskip \textrm{diag}\left(\mathbf{W}\right) = \mathbf{0}_n, W_{ij} = W_{ji} \geq 0, i \neq j,
    \end{split}
\end{equation}
where $\alpha, \beta > 0$ are tunable regularization hyperparameters. The logarithmic barrier imposed over the nodal degrees $\mathbf{d} = \mathbf{W}\mathbf{1}_n$ excludes isolated (null degree) vertices in the learned graph. The Frobenius-norm regularization on $\mathbf{W}$ controls edge sparsity via $\beta$.

Problem \eqref{eq:graph-learning-problem} can be further simplified. For instance, the free-decision variables can be arranged into a compact vector $\mathbf{w} \coloneqq \textrm{vec}\left(\textrm{triu}\left(\mathbf{W}\right)\right)\in \mathbb{R}_{+}^{r}$, $r = \frac{n(n-1)}{2}$, i.e. the upper-triangular elements of $\mathbf{W}$ (since it is symmetric and hollow). To impose $W_{ij}\geq 0$, we augment the cost with an indicator function $\iota_{\geq \mathbf{0}}\left(.\right)$, which takes the value of zero when its argument is a vector with non-negative entries, otherwise it is equal to $+\infty$. Given these definitions, one can rewrite \eqref{eq:graph-learning-problem} as an equivalent unconstrained, non-differentiable problem 
\begin{equation}
\label{eq:graph-learning-problem-vec}
\argminopt_{\mathbf{w} \in \mathbb{R}^{r}}\enskip 2\mathbf{z}^\top\mathbf{w} + \beta\left\|\mathbf{w}\right\|_{2}^{2} + \iota_{\geq \mathbf{0}}\left(\mathbf{w}\right) - \alpha\mathbf{1}_n^\top\log\left(\mathbf{S}\mathbf{w}\right),
\end{equation}
where $\mathbf{S} \in \left\{0,1\right\}^{n \times r}$ maps vectorized edge weights to nodal degrees.

\section{Proximal ADMM for Graph Learning}
\label{sec:ad-lpmm-for-graph-learning}

Problem \eqref{eq:graph-learning-problem-vec} has a structure amenable to ADMM-type solvers, as first recognized in~\cite{Wang2021}. Here we briefly review the batch PADMM iterations~\cite{Wang2023} that inspired our online algorithm in Section \ref{sec:online-graph-learning-using-ad-lpmm}.

\subsection{Applying ADMM to the graph learning problem}

Going back to \eqref{eq:graph-learning-problem-vec}, we introduce an auxiliary variable $\mathbf{v} \in \mathbb{R}^{n}$ (a proxy for $\mathbf{d}$) and the variable-splitting constraint $\mathbf{v} = \mathbf{S}\mathbf{w}$, to yield
\begin{equation}
\label{eq:split-problem}
\argminopt_{\mathbf{w} \in \mathbb{R}^{r},\mathbf{v} \in \mathbb{R}^{n}}\enskip f\left(\mathbf{w}\right) + g\left(\mathbf{v}\right) \enskip \text{subject to} \enskip \mathbf{S}\mathbf{w} - \mathbf{v} = \mathbf{0}_n,
\end{equation}
where the functions $f$ and $g$ are defined as
\begin{align*}
    f\left(\mathbf{w}\right) = {}& 2\mathbf{z}^\top\mathbf{w} + \beta\left\|\mathbf{w}\right\|_{2}^{2} + \iota_{\geq \mathbf{0}}\left(\mathbf{w}\right)\\
    g\left(\mathbf{v}\right) = {}& -\alpha\mathbf{1}_n^\top\log\left(\mathbf{v}\right).
\end{align*}
For $\rho>0$, the augmented Lagrangian of the problem \eqref{eq:split-problem} is given by
\begin{equation*}
\mathcal{L}_{\rho}\left(\mathbf{w},\mathbf{v},\pmb{\lambda}\right) = f\left(\mathbf{w}\right) + g\left(\mathbf{v}\right) + \pmb{\lambda}^\top\left(\mathbf{S}\mathbf{w}-\mathbf{v}\right) + \frac{\rho}{2}\left\|\mathbf{S}\mathbf{w}-\mathbf{v}\right\|_{2}^{2}.
\end{equation*}

PADMM generalizes ADMM by incorporating quadratic
proximity terms in the primal subproblems~\cite{Beck2017}. Given PSD matrices $\mathbf{G}$ and $\mathbf{H}$, the $k$-th PADMM iteration consists of three updates
\begin{align}
    \label{eq:w-update}
    \mathbf{w}^{(k+1)} = {}& \argminopt_{\mathbf{w} \in \mathbb{R}^{r}}\enskip\mathcal{L}_{\rho}(\mathbf{w},\mathbf{v}^{(k)},\pmb{\lambda}^{(k)}) + \|\mathbf{w}-\mathbf{w}^{(k)}\|_{\mathbf{G}/2}^{2},\\
    \label{eq:v-update}
    \mathbf{v}^{(k+1)} = {}& \argminopt_{\mathbf{v} \in \mathbb{R}^{n}}\enskip\mathcal{L}_{\rho}(\mathbf{w}^{(k+1)},\mathbf{v},\pmb{\lambda}^{(k)}) + \|\mathbf{v}-\mathbf{v}^{(k)}\|_{\mathbf{H}/2}^{2},\\
    \label{eq:lambda-update-1}
    \pmb{\lambda}^{(k+1)} = {}& \pmb{\lambda}^{(k)} + \rho(\mathbf{S}\mathbf{w}^{(k+1)} - \mathbf{v}^{(k+1)}).
\end{align}
%
Note that when $\mathbf{G} = \mathbf{0}_{r\times r}$ and $\mathbf{H} = \mathbf{0}_{n\times n}$, the proximity terms vanish and the updates \eqref{eq:w-update}-\eqref{eq:lambda-update-1} reduce to those of the standard ADMM. Typical choices for these matrices are $\mathbf{G} = \tau_{1}^{-1}\mathbf{I}_r - \rho\mathbf{S}^\top\mathbf{S}$ and $\mathbf{H} = (\tau_{2}^{-1} - \rho)\mathbf{I}_n$. Parameter values $0 < \tau_{1} < \frac{1}{\rho\left\|\mathbf{S}\right\|_{2}^{2}}$ and $0 < \tau_{2} < \frac{1}{\rho}$, with $\left\|\mathbf{S}\right\|_{2}^2=2(n-1)$~\cite{Saboksayr2021-accelerated} 
, guarantee the PSD requirement and convergence of the PADMM algorithm~\cite{Wang2023}.


\subsection{Efficient proximal updates}

The primal updates \eqref{eq:w-update} and \eqref{eq:v-update} can be expressed in terms of proximal operators, as
%
\begin{equation*}
\mathbf{w}^{(k+1)} = \proxopt_{\tau_{1}\cdot f}\left(\overline{\mathbf{w}}\right),\quad\quad
\mathbf{v}^{(k+1)} = \proxopt_{\tau_{2}\cdot g}\left(\overline{\mathbf{v}}\right),
\end{equation*}
where the auxiliary variables $\overline{\mathbf{w}}$ and $\overline{\mathbf{v}}$ are given by
\begin{align*}
\overline{\mathbf{w}} = {}& \mathbf{w}^{(k)}-\tau_{1}\cdot\left[\rho\cdot\mathbf{S}^\top\left(\mathbf{S}\mathbf{w}^{(k)}-\mathbf{v}^{(k)} + \frac{\pmb{\lambda}^{(k)}}{\rho}\right)\right],\\
\overline{\mathbf{v}} ={}& \mathbf{v}^{(k)}-\tau_{2}\cdot\rho\left(\mathbf{S}\mathbf{w}^{(k+1)}-\mathbf{v}^{(k)}-\frac{\pmb{\lambda}^{(k)}}{\rho}\right).
\end{align*}
To arrive at the custom PADMM algorithm for graph learning in~\cite{Wang2023}, we recognize both proximal operators admit simple expressions  
\begin{align*}
\proxopt_{\tau\cdot f}\left(\mathbf{w}\right) = {}&\max\left(\frac{\mathbf{w}-2\tau\mathbf{z}}{2\tau\beta+1},\mathbf{0}_r\right),\\
\proxopt_{\tau\cdot g}\left(\mathbf{v}\right) = {}&\frac{\mathbf{v} + \sqrt{\mathbf{v}^{2}+4\tau\alpha\mathbf{1}_n}}{2}.
\end{align*}
which act entrywise on their vector arguments. All in all, the PADMM updates at the $k$-th iteration become
\begin{align}
\label{eq:w-proximal-update}
\mathbf{w}^{(k+1)} = {}&  \max\left(\frac{\overline{\mathbf{w}}-2\tau_{1}\mathbf{z}}{2\tau_{1}\beta+1},\mathbf{0}_r\right),\\
\label{eq:v-proximal-update}
\mathbf{v}^{(k+1)} = {}&\frac{1}{2}\left(\overline{\mathbf{v}} + \sqrt{\overline{\mathbf{v}}^{2}+4\tau_{2}\alpha\mathbf{1}_n}\right),\\
\label{eq:lambda-update-2}
\pmb{\lambda}^{(k+1)} = {}&\pmb{\lambda}^{(k)} + \rho\left(\mathbf{S}\mathbf{w}^{(k+1)} - \mathbf{v}^{(k+1)}\right).
\end{align}
PADMM enjoys a global convergence rate of $\mathcal{O}(\frac{1}{k})$, and it has recently been shown to exhibit local convergence properties with a significantly faster rate of $\mathcal{O}(\mu^{k})$, $0 < \mu < 1$ \cite{Han2017}; see also~\cite{Wang2023}.

\section{Online Graph Learning from Streaming Signals}
\label{sec:online-graph-learning-using-ad-lpmm}

\subsection{Online PADMM algorithm}

PADMM's favorable convergence properties in the batch topology inference setting discussed so far, motivates well its adoption for online learning. Online estimation of $\mathbf{W}$ (or even tracking $\mathbf{W}^{(k)}$ in dynamic settings) using streaming signals $\{\mathbf{x}^{(1)},\dots,\mathbf{x}^{(k)},\mathbf{x}^{(k+1)},\dots\}$ can be cast as the following time-varying optimization problem
\begin{equation}
\label{eq:online-graph-learning-problem-vec}
\hat{\mathbf{w}}^{(k)} \in \argminopt_{\mathbf{w} \in \mathbb{R}^{n}}\enskip \overbrace{2\mathbf{z}_{1:k}^\top\mathbf{w} + \beta\left\|\mathbf{w}\right\|_{2}^{2} + \iota_{\geq \mathbf{0}}\left(\mathbf{w}\right)}^{f^{(k)}\left(\mathbf{w}\right)} \underbrace{- \alpha\mathbf{1}_n^\top\log\left(\mathbf{S}\mathbf{w}\right)}_{g\left(\mathbf{S}\mathbf{w}\right)}.
\end{equation}
Notice that the non-smooth, strongly convex component $f^{(k)}$ is time-varying because of the vectorized dissimilarity matrix $\mathbf{z}_{1:k}$, which is formed using all signals $\{\mathbf{x}^{(k)}\}$ acquired by time $k$; see also \eqref{eq:adjacency-similarity}. 

While \eqref{eq:online-graph-learning-problem-vec} can be solved by sequentially running a batch algorithm every time a new graph signal arrives, such procedure would incur in high delay and computational burden -- an unsatisfactory solution for delay-sensitive tasks. Furthermore, searching for a high-precision solution in dynamic network settings may not be worth the effort, since the new estimate $\hat{\mathbf{w}}^{(k+1)}$ may substantially deviate from the prior estimate $\hat{\mathbf{w}}^{(k)}$. Accordingly, we propose an online (recursive) algorithm to track the solution of the time-varying problem \eqref{eq:online-graph-learning-problem-vec}, which builds on the PADMM foundations in Section \ref{sec:ad-lpmm-for-graph-learning}.

Our online method consists on two steps per time instant $k=1,2,\ldots$. First, 
we recursively update the upper-triangular entries of the Euclidean distance matrix $\mathbf{z}_{1:k}$, once a new datum $\mathbf{x}^{(k)}$ is acquired at time $k$. Specifically, given $\mathbf{x}^{(k)}$ we form the vectorized dissimilarity matrix $\overline{\mathbf{z}}^{(k)}$ and update the running average
\begin{equation}
\label{eq:adjacency-similarity}
\mathbf{z}_{1:k} = \left(1-\gamma^{(k)}\right)\mathbf{z}_{1:k-1} + \gamma^{(k)}\overline{\mathbf{z}}^{(k)},
\end{equation}
where $\gamma^{(k)} \in (0,1)$ is a forgetting factor. In stationary settings (the graph is static), we choose $\gamma^{(k)} = \frac{1}{k}$, which renders \eqref{eq:adjacency-similarity} an infinite-memory running sample average. In dynamic environments we select $\gamma^{(k)} = 2 \times 10^{-3}$, so that \eqref{eq:adjacency-similarity} becomes an exponentially-weighted moving average with the ability to track topology changes. In the second step, our online algorithm updates the primal and dual vectors $\mathbf{w}^{(k)}$, $\mathbf{v}^{(k)}$ and $\pmb{\lambda}^{(k)}$ by running a single iteration of the batch PADMM updates \eqref{eq:w-proximal-update}, \eqref{eq:v-proximal-update}, \eqref{eq:lambda-update-2}, respectively; but employing $\mathbf{z}_{1:k}$ instead of $\mathbf{z}$ in \eqref{eq:w-proximal-update} to reflect the online nature of our method. The resulting topology estimate at time $k$ is stored in $\mathbf{w}^{(k)}$, and the overall online (O)PADMM procedure is tabulated under Algorithm \ref{alg:online-ad-lpmm-for-graph-learning}.
\begin{algorithm}[t]
\SetAlgoNoLine
\KwIn{Edge weight-to-node degree map $\mathbf{S}$; new dissimilarity information $\overline{\mathbf{z}}^{(k)}$ at time $k$ ; regularization hyperparameters $\alpha$ and $\beta$; PADMM hyperparameters $\rho$, $\tau_1$ and $\tau_2$; forgetting factor $\gamma$; initial values $\mathbf{w}^{(0)}$, $\mathbf{v}^{(0)}$ and $\pmb{\lambda}^{(0)}$}
\KwOut{Tracking solution available at the $k$-th iteration $\mathbf{w}^{(k)}$}
\For{$k = 1,2,\dots$}{
    $\text{Update}\enskip\gamma^{(k)}$\\
    $\mathbf{z}_{1:k} \leftarrow \left(1-\gamma^{(k)}\right)\mathbf{z}_{1:k-1}+\gamma^{(k)}\overline{\mathbf{z}}^{(k)}$\\
    $\overline{\mathbf{w}} \leftarrow \mathbf{w}^{(k-1)}-\tau_{1}\rho\mathbf{S}^\top\left(\mathbf{S}\mathbf{w}^{(k-1)}-\mathbf{v}^{(k-1)} + \frac{\pmb{\lambda}^{(k-1)}}{\rho}\right)$\;\\
    $\mathbf{w}^{(k)} \leftarrow \frac{1}{2\tau_{1}\beta+1}\max\left(\overline{\mathbf{w}}-2\tau_{1}\mathbf{z}_{1:k},\mathbf{0}_r\right)$\;\\
    $\overline{\mathbf{v}} \leftarrow \left(1+\rho\tau_{2}\right)\mathbf{v}^{(k-1)}-\rho\tau_{2}\mathbf{S}\mathbf{w}^{(k)} + \tau_{2}\pmb{\lambda}^{(k-1)}$\;\\
    $\mathbf{v}^{(k)} \leftarrow \frac{1}{2}\left(\overline{\mathbf{v}} + \sqrt{\overline{\mathbf{v}}^{2}+4\tau_{2}\alpha\mathbf{1}_n}\right)$\;\\
    $\pmb{\lambda}^{(k)} \leftarrow \pmb{\lambda}^{(k-1)} + \rho \left(\mathbf{S}\mathbf{w}^{(k)} - \mathbf{v}^{(k)}\right)$
}
\caption{OPADMM for graph learning}
\label{alg:online-ad-lpmm-for-graph-learning}
\end{algorithm}

\subsection{Discussion, computational complexity and regret analyses}

It is worth noting that matrix-vector multiplications with $\mathbf{S}$ effectively boil down to tabular searches and additions of the vector entries. All other operations in Algorithm \ref{alg:online-ad-lpmm-for-graph-learning} are pointwise, so the computational cost per iteration is $\mathcal{O}\left(r\right)$ (recall $r = \frac{n(n-1)}{2}$). This is on par with online PG~\cite{Saboksayr2021} and DPG~\cite{Saboksayr2023}, and less expensive than the prediction-correction algorithm in~\cite{natali2022online}. 
Unlike batch algorithms for time-varying graph learning~\cite{Wang2023,vinicius2020asilomar}, OPADMM's memory storage and computational cost do not depend on the length of the data acquisition horizon. These batch algorithms augment the objective function \eqref{eq:graph-learning-problem} with a temporal-variation regularization, to enforce similarity between consecutive adjacency matrix estimates. Interestingly, this feature is seamlessly embedded in OPADMM by virtue of the proximity term in \eqref{eq:w-update} -- a significant improvement over the state-of-the-art online DPG algorithm.

To obtain performance guarantees in terms of static regret, we will henceforth assume that nodal degrees are in a compact set that is uniformly bounded away from zero (meaning $d_{\min}\mathbf{1}_n \preceq \mathbf{d} \preceq d_{\max}\mathbf{1}_n$, for some $0<d_{\min}\leq d_{\max}$). This is a loose requirement that ensures the subgradient of the objective has bounded norm. In practice if degrees become arbitrarily small, it is prudent to apply a threshold and remove the loosely connected nodes from $\mathcal{G}(\mathcal{V},\mathcal{E},\mathbf{W})$. With this extra requirement one can show that for $\tau_2=\rho$ so that $\mathbf{H}=\mathbf{0}_{n\times n}$, \cite[Theorem 3]{Wang2012} ensures that OPADMM attains a sublinear static regret, namely
\begin{equation*}
\sum_{k=1}^p f^{(k)}(\mathbf{w}^{(k)})+g(\mathbf{v}^{(k)})-\min_{\mathbf{S}\mathbf{w}=\mathbf{v}}\sum_{k=1}^p f^{(k)}(\mathbf{w})+g(\mathbf{v})\leq \mathcal{O}(\sqrt{p}).
\end{equation*}
The extension to the case where $\mathbf{H}\neq \mathbf{0}_{n\times n}$ is left as future work.

\begin{figure*}[ht]
  \centering
  \begin{tabular}{ccc}
  \subf{\includegraphics[width=5.4cm]{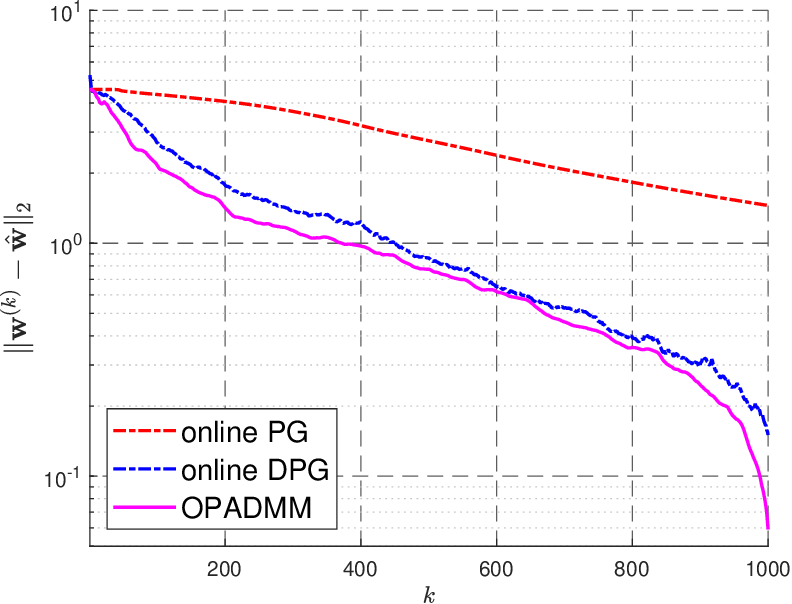}}{(a) Gaussian (stationary)} &
  \subf{\includegraphics[width=5.4cm]{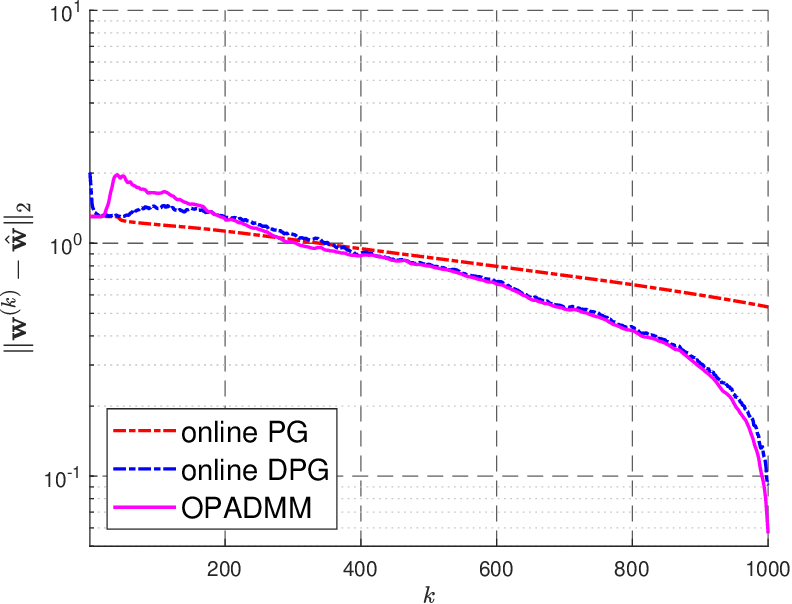}}{(b) Erd\"os-Rényi (stationary)} &
  \subf{\includegraphics[width=5.4cm]{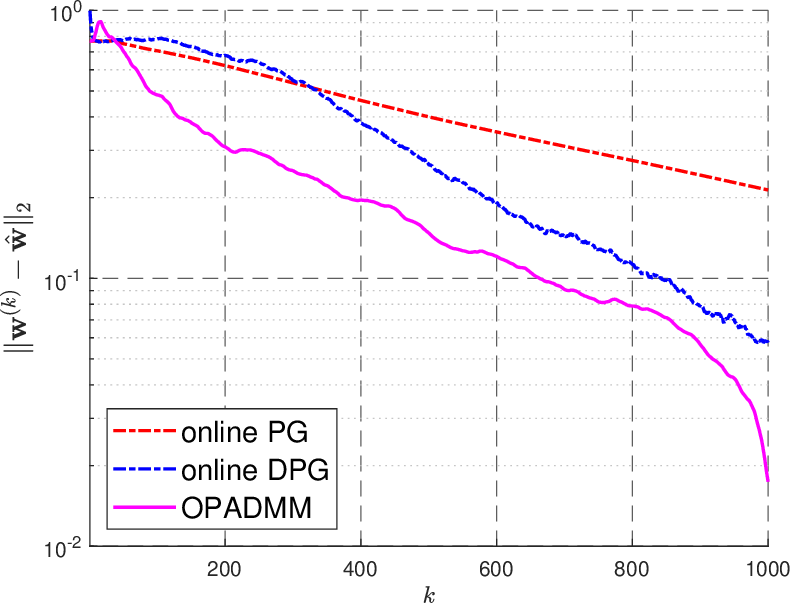}}{(c) Preferential attachment (stationary)}\\
  \subf{\includegraphics[width=5.4cm]{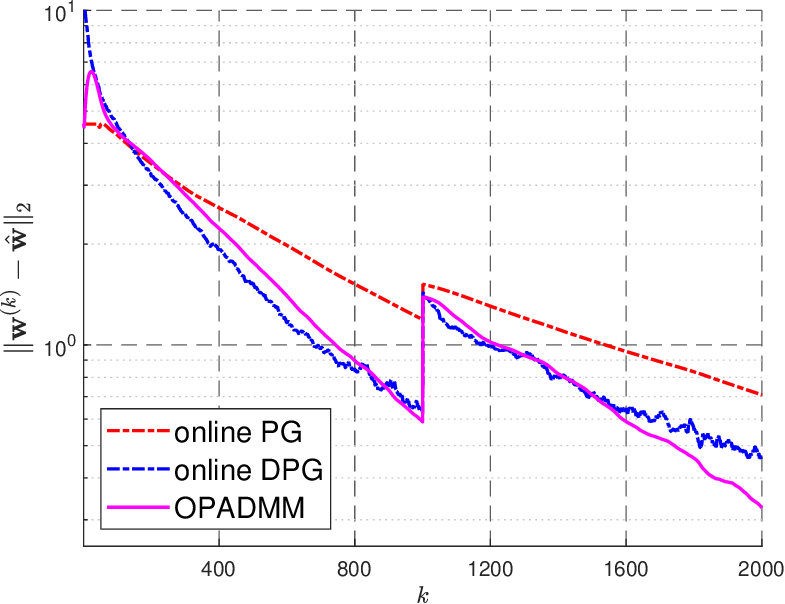}}{(d) Gaussian (dynamic)} &
  \subf{\includegraphics[width=5.4cm]{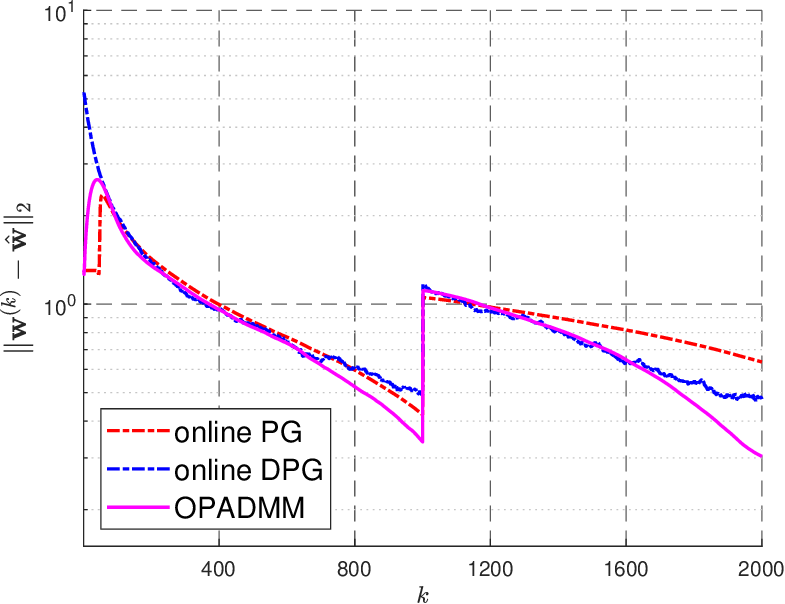}}{(e) Erd\"os-Rényi (dynamic)} &
  \subf{\includegraphics[width=5.4cm]{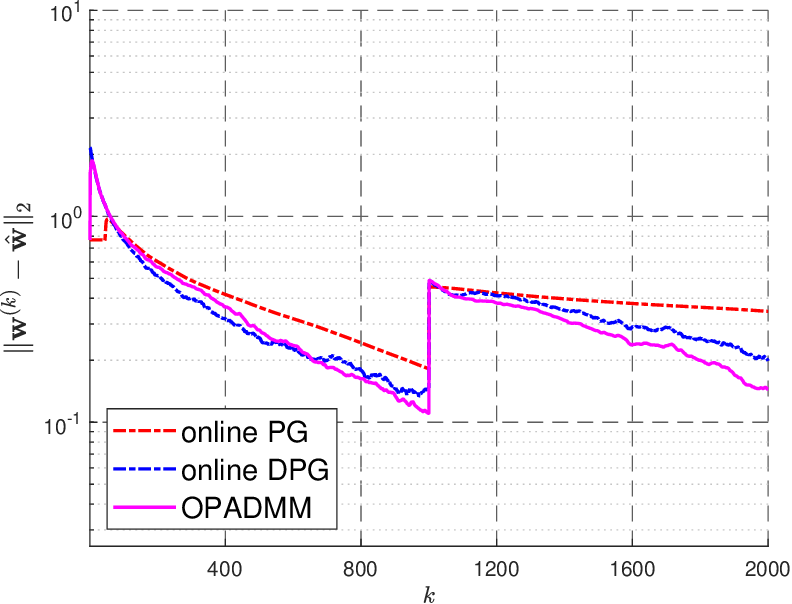}}{(f) Preferential attachment (dynamic)}
  \end{tabular}
  \caption{Convergence behavior illustrated via the evolution of suboptimality $\|\mathbf{w}^{(k)}-\hat{\mathbf{w}}\|_{2}$ for synthetic random graph models with $n=100$ nodes: (a)-(c) stationary graphs, and (d)-(f) time-varying graphs. For the dynamic networks, the topology changes at $k=1000$. In stationary settings, our proposed method converges faster than online PG and DPG algorithms, while in dynamic settings, it demonstrates superior adaptability to changes in network topology.}
  \label{fig:results-simulated-graphs}
\end{figure*}

\begin{figure*}[h]
  \centering
  \begin{tabular}{ccc}
  \subf{\includegraphics[width=5.4cm]{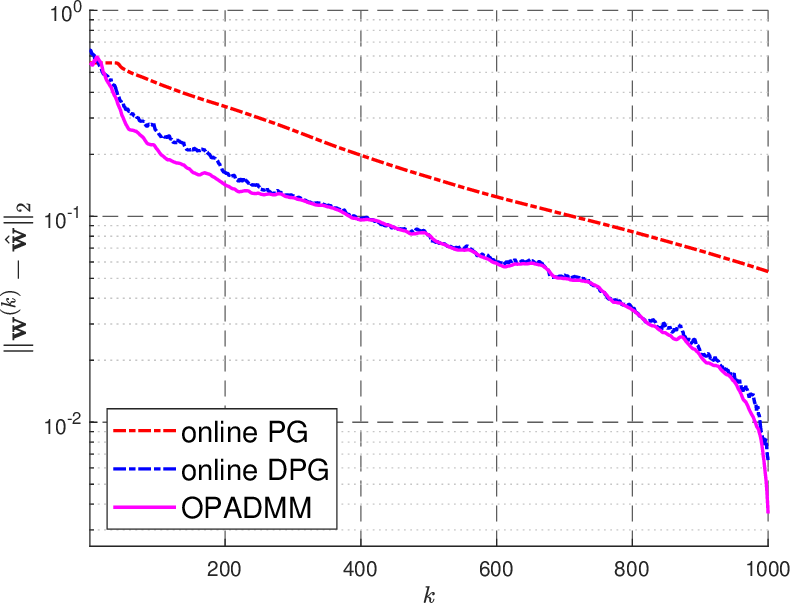}}{(a) mesh1e1} &
  \subf{\includegraphics[width=5.4cm]{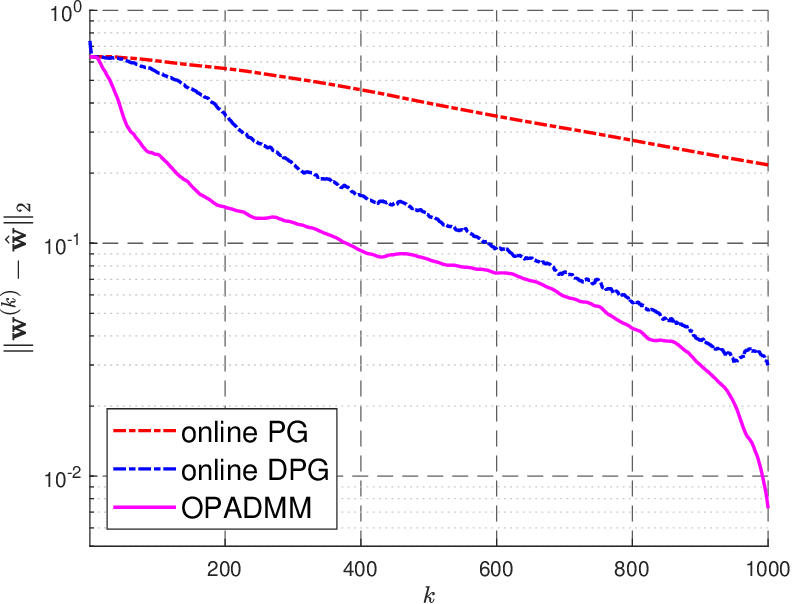}}{(b) bcspwr03} &
  \subf{\includegraphics[width=5.4cm]{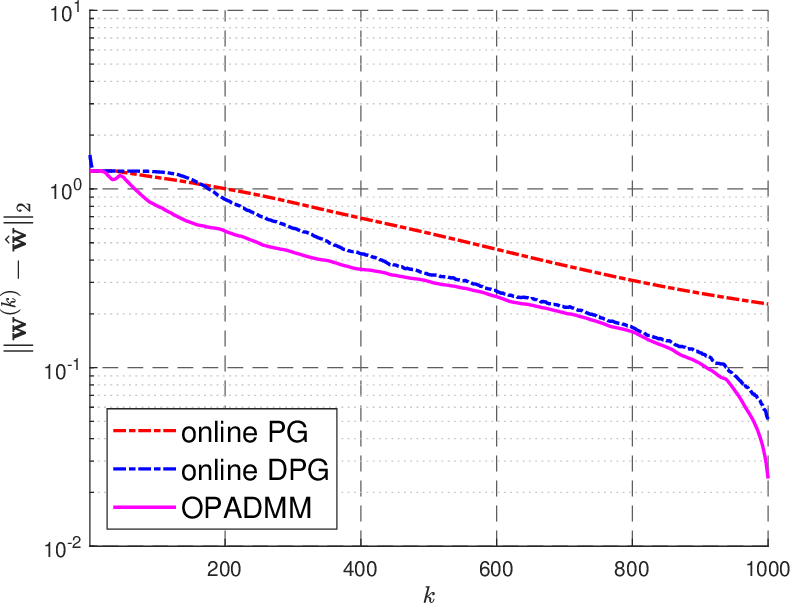}}{(c) lshp\_265}
  \end{tabular}
  \caption{Convergence behavior illustrated via the evolution of suboptimality $\|\mathbf{w}^{(k)}-\hat{\mathbf{w}}\|_{2}$ for real-world graphs from the datasets~\cite{Davis2011}. Our proposed method converges faster than online PG and DPG algorithms in all cases.}
  \label{fig:results-real-world-data}
\end{figure*}


\section{Experimental Evaluation}
\label{sec:computational-experiments}

Here we first test OPADMM using various graphs ensembles that were synthetically generated using different models. We evaluate the convergence of our proposed method by montoring the suboptimality (tracking error) $\|\mathbf{w}^{(k)}-\hat{\mathbf{w}}\|_{2}$, where $\mathbf{w}^{(k)}$ is the solution produced by Algorithm \ref{alg:online-ad-lpmm-for-graph-learning} at the $k$-th iteration, and $\hat{\mathbf{w}}$ corresponds to the solution of the batch method using the values of hyperparameters $\alpha$ and $\beta$ that are the best in a grid search evaluation using the method in \cite{Kalofolias2016}. Finally, we perform a grid search for the hyperparameters $\rho$, $\tau_{1}$ and $\tau_{2}$ of OPADMM and compare the best case with the state-of-the-art-methods online PG and DPG. We also use real world datasets to validate our proposed method in a similar fashion to the tests with synthetic graphs. The code to generate all figures in this work is available in a public GitHub repository at \url{https://github.com/hchahuara/ogl}.

\subsection{Synthetic graphs}

\noindent \textbf{Stationary graphs.} A set of $p=1000$ smooth graph signals, each one with $n=100$ entries, were synthetically generated and corrupted with Gaussian noise ($\mu = 0$, $\sigma^{2} = 0.01$). Graphs realizations were drawn from three random graph models: random geometric with Gaussian kernel, Erd\"os-Rényi (ER), and preferential attachment (PA). Sparse connections were achieved with threshold $0.8$ and scale $0.2$ for the Gaussian model, edge probability $0.1$ for the ER model, and $2$ connected nodes initially and then adding new nodes one at a time for the PA model. Numerical results for the aforementioned test cases are depicted in Figure \ref{fig:results-simulated-graphs} (a)-(c). For stationary graphs, OPADMM outperforms online PG and DPG in terms of convergence speed.\vspace{3pt}

\noindent \textbf{Time-varying graphs.} We generated another set of $p=2000$ smooth graph signals, each with $n=100$ elements, and similarly corrupted them with Gaussian noise ($\mu = 0$, $\sigma^{2} = 0.01$). These graph signals were designed to be smooth on time-varying (piecewise-stationary) graphs, with $10\%$ of the edges resampled after 1000 samples. Dynamic graphs were generated using the models with the same parameter values used for stationary graphs. The results are shown in Figure \ref{fig:results-simulated-graphs} (d)-(f), from where it is apparent that OPADMM outperforms PG in convergence. While online DPG initially tracks the solution marginally faster, ultimately, OPADMM adapts better to the network changes and ends up converging faster than DPG.

\subsection{Real world graphs}

We also test the performance of our proposed algorithm using three datasets from the SuiteSparse Matrix Collection \cite{Davis2011}. In particular, we use the structural engineering data mesh1e1 ($n=48$ nodes), power network data bcspwr03 ($n=118$ nodes), and thermal network data lshp\_265 ($n=265$ nodes). All experiments with these real graphs have been carried out by generating $p=1000$ synthetic smooth signals corrupted with Gaussian noise ($\mu = 0$, $\sigma^{2} = 0.01$). We report these results in Figure \ref{fig:results-real-world-data}. It can be observed that OPADMM clearly yields a faster convergence behavior in comparison with online PG and DPG in all the performed tests. Our findings are thus fairly robust.

\section{Conclusions}
\label{sec:conclusions}

In this paper, we have developed an efficient online optimization method for graph learning. Leveraging the local linear convergence properties of the PADMM and the structure of the network topology inference problem, we derive an alternating minimization algorithm (termed OPADMM) to minimize a time-varying cost in an online fashion. The proximity term in the topology updates implements a temporal-variation regularization, and we show OPADMM exhibits sublinear static regret under simplifying assumptions. Computational experiments demonstrate the effectiveness of OPADMM in stationary or dynamic settings, and that it outperforms state-of-the-art online algorithms for synthetically generated instances. Furthermore, the proposed approach exhibits robust performance across a broad range of graphs, as evidenced via tests using real world datasets.

\bibliographystyle{IEEEtran}
\bibliography{references}

\end{document}